\documentclass[12pt,aps,superscriptaddress,prl]{revtex4}

\usepackage{amsmath, amssymb}
\usepackage{xcolor, graphicx, ulem}
\usepackage[colorlinks=true,urlcolor=blue,citecolor=blue,linkcolor=black]{hyperref}
\usepackage[shortlabels]{enumitem}
\usepackage{amsthm}
\usepackage{amsfonts}

\usepackage{bbold}
\usepackage{stmaryrd}
\usepackage{color}
\usepackage{hyperref}
\usepackage{verbatim}
\usepackage{cases}
\usepackage{amssymb}
\usepackage[]{mathrsfs}
\usepackage{anysize}
\usepackage{epsfig}
\usepackage{bm}
\usepackage{setspace}
\usepackage{enumerate}
\usepackage{dcolumn}
\usepackage{bm}
\usepackage{enumitem}

\newcommand{\bra}[1]{\left< #1 \right\vert}
\newcommand{\ket}[1]{\left\vert #1 \right>}

\newcommand{\pare}[1]{\left( #1 \right)}

\newcommand{\ave}[1]{\left\langle #1 \right\rangle}
\newcommand{\im}{\mathrm{i}}

\newcommand{\PP}{\Psi}
\newcommand{\g}{\gamma}
\newcommand{\cc}{\kappa}

\renewcommand{\epsilon}{\varepsilon}

\renewcommand{\d}{\mathrm{d}}

\begin{document}

\title{Two-particle four-point correlations in dynamically disordered tight-binding networks}


\author{Armando Perez-Leija}\email{apleija@gmail.com}
\affiliation{Humboldt-Universit\"at zu Berlin, Institut f\"ur Physik, AG Theoretische Optik $\&$ Photonik, 12489 Berlin, Germany}
\affiliation{Max-Born-Institut, 12489 Berlin, Germany}
\author{Roberto de J. Leon-Montiel}
\affiliation{Instituto de Ciencias Nucleares, Universidad Nacional Aut\'onoma de M\'exico, 70-543, 04510 Cd. Mx., M\'exico}
\author{Jan Sperling}
\affiliation{Clarendon Laboratory, University of Oxford, Parks Road, Oxford OX1 3PU, United Kingdom}
\author{Hector Moya-Cessa}
\affiliation{Instituto Nacional de Astrof\'isica, \'Optica y Electr\'onica, Calle Luis Enrique Erro 1, Santa Mar\'ia
Tonantzintla, Puebla CP 72840, M\'exico}
\affiliation{Institut f\"{u}r Quantenphysik and Center for Integrated Quantum Science and Technology (IQST), Universit\"{a}t Ulm, Albert-Einstein-Allee 11, D-89081 Ulm, Germany}
\author{Alexander Szameit}
\affiliation{Institut f\"ur Physik, Universit\"at Rostock, D-18051 Rostock, Germany }
\author{Kurt Busch}
\affiliation{Humboldt-Universit\"at zu Berlin, Institut f\"ur Physik, AG Theoretische Optik $\&$ Photonik, 12489 Berlin, Germany}
\affiliation{Max-Born-Institut, 12489 Berlin, Germany}

\begin{abstract}
\begin{center}
\textbf{ABSTRACT}
\end{center}
We use the concept of two-particle probability amplitude to derive the stochastic evolution equation for two-particle four-point correlations in tight-binding networks affected by diagonal dynamic disorder. It is predicted that in the presence of dynamic disorder, the average spatial wave function of indistinguishable particle pairs delocalizes and populates all network sites including those which are weakly coupled in the absence of disorder. Interestingly, our findings reveal  that correlation elements accounting for particle indistinguishability are immune to the impact of dynamic disorder.
\end{abstract}

\maketitle
\newpage
\section{I. Introduction}
Non-classical correlations among many particles or quantized fields are key elements to understand and ultimately apply the nonlocal properties of quantum systems \cite{Aspect1982, Haroche2013, Wineland2013}. In this regard, of particular interest has been the study of quantum correlations between indistinguishable particles co-propagating in time-independent disordered lattices \cite{Lahini2010, Abouraddy2012, Giuseppe2013, Gilead2015}. Yet, most investigations have solely considered second-order two-particle correlations in lattices that are affected by static disorder in either the site energies or the coupling coefficients \cite{Abouraddy2012, Giuseppe2013, Gilead2015}. Interestingly, little has been said about the impact of dynamic disorder on the evolution of two-particle correlations in such lattice systems. Indeed, studies of dynamically disordered systems have focused on the evolution of single-excitations, a case in which the dynamics does not show any divergence from classical wave mechanics \cite{Yu2009, Roberto2015}. In the single-excitation subspace, a number of investigations have brought to light intriguing noise-induced phenomena. Among them one may mention environment-assisted quantum transport \cite{Plenio2008, Rebentrost2009, Eisfeld2012, Viciani2015, Leon2015, Schonleber2015, Biggerstaff2016, Caruso2016}, the emergence of virtual amplifying electric circuit elements \cite{Quiroz2016}, and the enhancement of particle transport through symmetric optical potentials \cite{Leon2017}.\\ 
In this contribution we investigate the dynamics of two-particle correlation functions in which the associated two-particle probability amplitudes are evaluated at four different points of a network exhibiting dynamic disorder in the site energies. Conventionally, transverse disorder in the site energies of coupled systems is termed diagonal disorder \cite{Lane2011}. Hence, the systems analyzed here are networks affected by dynamic diagonal disorder.\\
This work is organized as follows. In section II, we introduce the concept of two-particle probability amplitude and derive the evolution equation for two-particle four-point correlation functions in finite disorder-free networks. In section III, we proceed to develop the evolution equation for two-particle four-point correlation functions propagating in tight-binding networks affected by dynamic disorder. Finally, we conclude in section IV. 

\section{II. Two-particle probability amplitude in tight-binding networks}
In this section we develop the concept of two-particle probability amplitude and derive the corresponding equations of motion for finite tight-binding networks comprising $N$ sites. To do so, we first note that the probability amplitudes for a quantum particle, initialized at site $n$, within a tight-binding network are governed by the equations
\begin{equation} \label{eq:1}
\im\frac{\d U_{p,n}}{\d t}=-\epsilon_{p} U_{p,n}-\sum_{r=1}^{N}\cc_{p,r}U_{r,n}.
\end{equation}
Here, we set $\hbar=1$, $\epsilon_{p}$ represents the energy at the $p$-th site, and $\cc_{p,r}$ are the coupling coefficients between sites $p$ and $r$. In terms of single-particle probability amplitudes, $U_{p,n}$, one can define the two-particle probability amplitudes at sites $p$ and $q$ as
\begin{equation}\label{eq:2}
\begin{aligned}
\PP_{p,q}(t)=\sum_{m=1,n=1}^{N,N}\varphi_{m,n}\left[U_{p,n}(t)U_{q,m}(t)\pm U_{p,m}(t)U_{q,n}(t)\right],
\end{aligned}
\end{equation}
where $\varphi_{m,n}$ is the initial probability amplitude profile which fulfills the condition $\sum_{m=1,n=1}^{N,N}|\varphi_{m,n}|^{2}=1$. Additionally, the $\pm$ sign determines whether the particles are bosons $(+)$ or fermions $(-)$, respectively. By taking the time derivative of $\PP_{p,q}(t)$ and using Eq.~\eqref{eq:1} we obtain the two-particle evolution equation
\begin{equation}\label{eq:3}
\begin{split}
\frac{\d\PP_{p,q}}{\d t}=\im\left(\epsilon_{p}+\epsilon_{q}\right)\PP_{p,q}+\im\sum_{r}\left( \cc_{p,r}\PP_{r,q} + \cc_{q,r}\PP_{p,r}\right).
\end{split}
\end{equation}
\begin{figure}[t!]
\includegraphics[width=\linewidth]{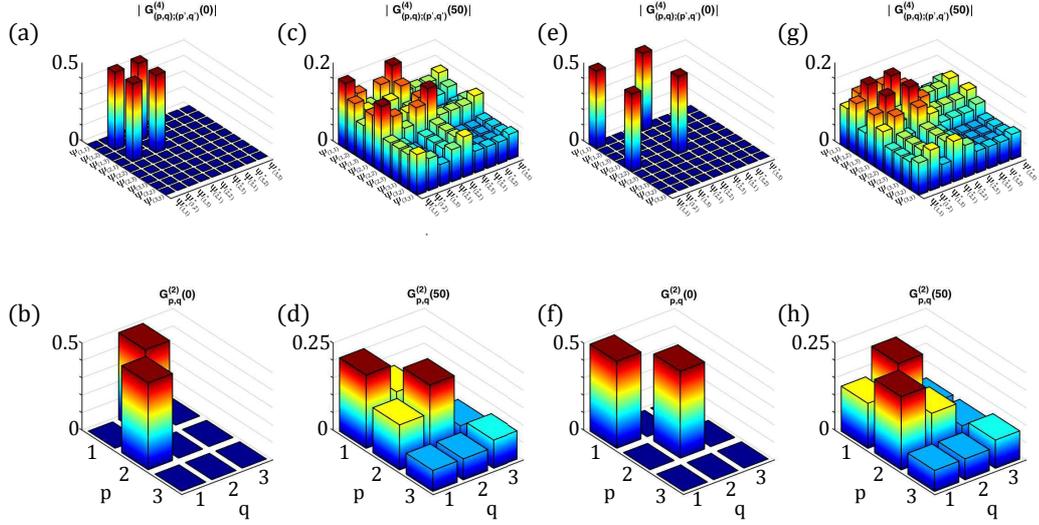}
\caption{(Top row) Four-point correlation matrices for a two-particle separable state (a) and an entangled state (e). In (c) and (g) it is shown the evolved correlation maps at $t=50$ $\left(G^{(4)}_{(p,q);(p',q')}(50)\right)$. (Bottom row) Two-point correlation matrices obtained from the diagonal elements of (a), (c), (e), and (g), respectively.}
\label{Fig1}
\end{figure}
Accordingly, two-particle quantum states evolve in a Hilbert space composed of a discrete set of $N^{2}$-mode states occupied by a total of exactly two particles. One significant fact to emphasize regarding Eq.~\eqref{eq:3} is the presence of the term $\left(\epsilon_{p}+\epsilon_{q}\right)\PP_{p,q}$, which implies that during the evolution $\PP_{p,q}$ acquires a phase that a single particle acquires when it traverses the same network twice \cite{Giuseppe2013}. Indeed, such effects can be expected since we are dealing with two particles \cite{Klyshko1982}. An important aspect associated with $\PP_{p,q}$ is that its modulus squared provides the two-point correlation function $G^{(2)}_{p,q}(t)=|\PP_{p,q}(t)|^{2}$, which gives the probability of finding one particle at site $p$ and the other at $q$ \cite{Abouraddy2001, Saleh2005, Abouraddy2002, Bromberg2009, Lebugle2015, Weimann2016}. 

Moreover, since the two-point correlation function can be expressed as $G^{(2)}_{p,q}(t)=\PP_{p,q}(t)\PP_{p,q}^{*}(t)$, it is natural to define the four-point correlation function as $G^{(4)}_{(p,q);(p',q')}(t)=\PP_{p,q}(t)\PP_{p',q'}^{*}(t)$ \cite{Loudon2000}, whose time evolution is governed by the equation
\begin{equation}\label{eq:4}
\begin{split}
&\frac{\d}{\d t}G^{(4)}_{(p,q);(p',q')}=\im(\epsilon_{p}+\epsilon_{q}-\epsilon_{p'}-\epsilon_{q'})G^{(4)}_{(p,q);(p',q')}\\
&+\im\sum_{r}\left(\kappa_{p,r}G^{(4)}_{(r,q);(p',q')}+\kappa_{q,r}G^{(4)}_{(p,r);(p',q')}\right)\\
&-\im\sum_{r'}\left(\kappa_{p',r'}G^{(4)}_{(p,q);(r',q')}+\kappa_{q',r'}G^{(4)}_{(p,q);(p',r')}\right).
\end{split}
\end{equation}
\begin{figure}[t!]
\includegraphics[width=\linewidth]{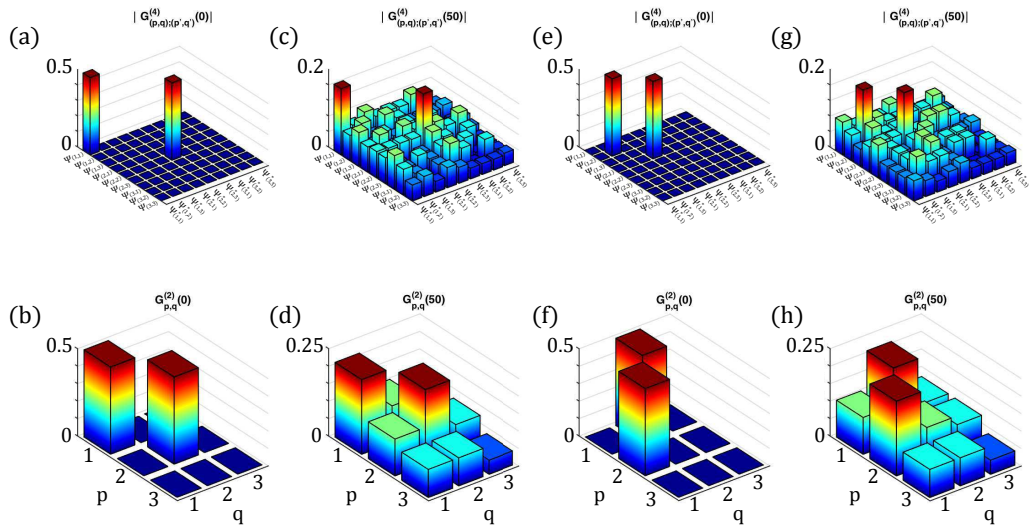}
\caption{(Top row) Four-point correlation matrices for a two-particle classically correlated state (a) and an incoherent state (e). In (c) and (g) it is shown the evolved correlation maps at $t=50$ $\left(G^{(4)}_{(p,q);(p',q')}(50)\right)$. (Bottom row) Two-point correlation matrices obtained from the diagonal elements of (a), (c), (e), and (g), respectively.}
\label{Fig2}
\end{figure}
The derivation of this expression is straightforward by computing $\frac{\d}{\d t}\left(\PP_{p,q}(t)\PP_{p',q'}^{*}(t)\right)$ and using Eq.~\eqref{eq:3} and its complex conjugate. Before considering particular examples, it is worth noting that $\PP_{p,q}(t)\PP_{p',q'}^{*}(t)$ describes the coherence between the states $\ket{1_{p},1_{q}}$ and $\ket{1_{p'},1_{q'}}$, consequently $G^{(4)}_{(p,q);(p',q')}(t)$ can be thought of as the two-particle density matrix.\\
For illustrative purposes, we examine two-particle four-point correlations that occur in a network consisting of 3 sites with energies $\epsilon_{1}=\epsilon_{2}=\epsilon_{3}=1$, and coupling coefficients $\cc_{1,2}=1$, $\cc_{1,3}=1/3$, $\cc_{2,3}=1/3$. For simplicity, $\epsilon_{n}$ and $\kappa_{m,n}$ are given in normalized units. 
As initial states we consider four bosonic cases: i) two particles in the separable state $\ket{\PP(0)}=\frac{1}{\sqrt{2}}\left(\ket{1_{1},1_{2}}+\ket{1_{2},1_{1}}\right)$, ii) two particles in the entangled state $\ket{\PP(0)}=\frac{1}{\sqrt{2}}\left(\ket{1_{1},1_{1}}+\ket{1_{2},1_{2}}\right)$, iii) two \textit{classically correlated} particles described by the density matrix $\rho(0)=\frac{1}{2}\left(\ket{1_{1},1_{1}}\bra{1_{1},1_{1}}+\ket{1_{2},1_{2}}\bra{1_{2},1_{2}}\right)$, and iv) two distinguishable particles represented by $\rho(0)=\frac{1}{2}\left(\ket{1_{1},1_{2}}\bra{1_{1},1_{2}}+\ket{1_{2},1_{1}}\bra{1_{2},1_{1}}\right)$. Throughout this work we use the compact notation $\ket{1_{m},1_{n}}$ to represent the state $\ket{1_{m}}\otimes\ket{1_{n}}$, and it represents a state where one particle is at site $m$ and another at $n$. Additionally, states $\propto\left(\ket{1_{m},1_{n}}+\ket{1_{n},1_{m}}\right)$, are symmetrized wavefunctions. 
Notice that in i), ii), and iii) the particles are assumed to be indistinguishable, while in iv) they are distinguishable. Additionally, we point out that the classically correlated state represents a superposition of particle-pair probabilities at sites $1$ and $2$, while the entangled state is a superposition of probability amplitudes. The reason for exploring the dynamics of mixed states iii) and iv) is discussed in the next section. Figs.~(\ref{Fig1}.a, e) and (\ref{Fig2}.a, e) (Figs.~(\ref{Fig1}.b, f) and (\ref{Fig2}.b, f)) depict the four-point (two-point) correlation matrices for the initial states. \\ 
Once the particles evolve into the system, $G^{(4)}_{(p,q);(p',q')}$ reveals the coherent superpositions shown in Figs.~(\ref{Fig1}) and (\ref{Fig2}), where the specific time $t=50$ was randomly chosen. For the input states i) and ii), the computed spatial coherences are presented in Figs.~(\ref{Fig1}.c) and~(\ref{Fig1}.g), respectively. Evidently, since the coupling between the sites one and two is three times larger than the coupling between the sites one and three and two and three,  $\PP_{p,q}$ propagates in a coherent fashion hopping predominantly among the strongly-coupled sites, and the lower correlation amplitudes are registered in site-pairs $(p,q; p',q')$ involving the third (lowest-coupled) site. By looking at the $G^{(2)}_{(p,q)}$ for separable, Fig.~(\ref{Fig1}.d), and entangled, Fig.~(\ref{Fig1}.h), input states, it is clear that bunching and anti-bunching effects are dominant, respectively. However, the $G^{(4)}_{(p,q);(p',q')}$ in Figs.~(\ref{Fig1}.c) and (\ref{Fig1}.g), reveal that the coherences $\PP_{1,1}\PP^{*}_{2,2}$ and $\PP_{1,2}\PP^{*}_{2,1}$ are equally likely to occur, respectively. In both cases, the coherences $\PP_{2,1}\PP^{*}_{1,1}$ and $\PP_{2,2}\PP^{*}_{1,2}$ are the second higher elements.
For the classically correlated [Figs.~(\ref{Fig2}.a, c)] and the incoherent case  [Figs.~(\ref{Fig2}.e, g)], the highest correlation peaks are the incoherent terms $\left(\mid \PP_{1,1}\mid^{2}, \mid\PP_{2,2}\mid^{2}\right)$  and $\left(\mid\PP_{1,2}\mid^{2}, \mid\PP_{2,1}\mid^{2}\right)$, respectively. 
\section{III. Two-particle four-point correlations in dynamically disordered networks}
In order to obtain the equations of motion for two-particle states traversing dynamically disordered networks, we start by writing the stochastic Schr\"odinger equation for the matrix elements of the single-particle evolution operator
\begin{equation}\label{eq:5}
\frac{\d U_{q,n}}{\d t} =\im\epsilon_{q}\pare{t}U_{q,n}+\im\sum_{r}\kappa_{r,q}U_{r,n}, 
\end{equation}
where $\epsilon_{q}\pare{t}=\epsilon_{q} + \phi_{q}\pare{t}$. Here $\epsilon_{q}$ represents the average energy of the $q$-th site and $\phi_{q}\pare{t}$ describes a random Gauss-Markov process with zero average (Wiener process) \cite{Jacobs_book}, i.e., $\ave{\phi_{q}\pare{t}}=0$ and $\ave{\phi_{q}\pare{t}\phi_{p}\pare{t'}}=\gamma_{q}\delta_{qp}\delta\pare{t-t'}$, where $\gamma_{q}$ stands for the noise intensity (dephasing rate) and $\ave{\cdots}$ denotes stochastic averaging.
Due to the stochastic nature of the site energies, we rely on stochastic calculus to derive the evolution equation governing four-point correlations. 
As shown in the Appendix, within the It\^{o}'s calculus framework \cite{vanKampen1981}, Eqs.~\eqref{eq:5} and \eqref{eq:2} yield the differential of $\PP_{p,q}$ 
\begin{equation}\label{eq:diff_Psi}
\begin{split}
\d\PP_{p,q}=&\left[\im\left(\epsilon_{q}+\epsilon_{p}\right)\PP_{p,q}+\im\sum_{r}\cc_{r,q}\PP_{p,r}\right. \\ & \left. +\im\sum_{r}\cc_{r,p}\PP_{r,q}-\frac{1}{2}\left(\g_{p}+\g_{q}\right)\PP_{p,q}\right]\d t\\
& +\im\sqrt{\g_{q}}\PP_{p,q}\d W_{q} +\im\sqrt{\g_{p}}\PP_{p,q}\d W_{p}  \\
& -\sqrt{\g_{p}\g_{q}}\PP_{p,q}\d W_{q}\d W_{p},
\end{split}
\end{equation}
where we have introduced the Wiener increments $\d W_{p}=\phi_{p}(t)\d t/\sqrt{\g_{p}}$ that fulfill the conditions $\ave{\d W_{q,p}}=0$ and $\ave{\d W_{q}\d W_{p}}=\delta_{qp}\d t$ \cite{Jacobs_book}. Using Eq. (\ref{eq:diff_Psi}) along with the It\^{o}'s product rule, $\d\left(\PP_{p,q}\PP^{*}_{p',q'}\right)=\d\left(\PP_{p,q}\right)\PP^{*}_{p',q'}+\PP_{p,q}\d\left(\PP^{*}_{p',q'}\right)+\d\left(\PP_{p,q}\right)\d\left(\PP^{*}_{p',q'}\right)$, we obtain the evolution equation for the mean four-point correlation function $\tilde{G}^{(4)}_{(p,q);(p',q')} = \left\langle \PP_{p,q}\PP_{p',q'}^{*}\right\rangle$ as
\begin{widetext}
\begin{eqnarray} \label{eq:average}
\frac{\d}{\d t} \tilde{G}^{(4)}_{(p,q);(p',q')}
& = & \Bigg[\im\left(\epsilon_{p}+\epsilon_{q}-\epsilon_{p'}-\epsilon_{q'}\right)-\frac{1}{2}\left(\g_{p}+\g_{q}+\g_{p'}+\g_{q'}\right) +\sqrt{\g_{p}\g_{p'}}\delta_{p,p'}+\sqrt{\g_{q}\g_{q'}}\delta_{q,q'} \nonumber \\
&  &\;\;\; \left. +\sqrt{\g_{p}\g_{q'}}\delta_{p,q'}+\sqrt{\g_{q}\g_{p'}}\delta_{q,p'}\right. -\sqrt{\g_{p}\g_{q}}\delta_{p,q}-\sqrt{\g_{p'}\g_{q'}}\delta_{p',q'}\Bigg] \tilde{G}^{(4)}_{(p,q);(p',q')}\nonumber \\
 &  & + \im\Bigg[\sum_{r}\cc_{r,q} \tilde{G}^{(4)}_{(p,r);(p',q')}+\sum_{r}\cc_{r,p}\tilde{G}^{(4)}_{(r,q);(p',q')} -\sum_{r'}\cc_{r',q'}  \tilde{G}^{(4)}_{(p,q);(p',r')}\nonumber \\
 &  & \;\;\;  -\sum_{r'}\cc_{r',p'} \tilde{G}^{(4)}_{(p,q);(r',q')}\Bigg].
\end{eqnarray}
\end{widetext}
It should be noted that the first term on the right-hand side of Eq.~\eqref{eq:average} is complex and it vanishes for all diagonal elements $ \tilde{G}^{(4)}_{(p,q);(p,q)}$ as well as for the off-diagonal elements accounting for particle indistinguishability $\tilde{G}^{(4)}_{(p,q);(q,p)}$ (that is when $(p, q)=(p', q')$ or $(p, q)=(q, p)$ the first term becomes zero). On the other hand, for the remaining off-diagonal elements $\tilde{G}^{(4)}_{(p,q);(p',q')}$, such a term becomes $\left(-\epsilon_{p}-\epsilon_{q}+\epsilon_{p'}+\epsilon_{q'}\right)-\frac{\im}{2}\left(\g_{p}+\g_{q}+\g_{p'}+\g_{q'}\right)$.
Owing to the negativity of the imaginary part, we determine that the off-diagonal elements, $\tilde{G}^{(4)}_{(p,q);(p',q')}$, will vanish as they are affected by an attenuation factor arising from noise intensities.\\ 
\begin{figure}[t!]
\includegraphics[width=\linewidth]{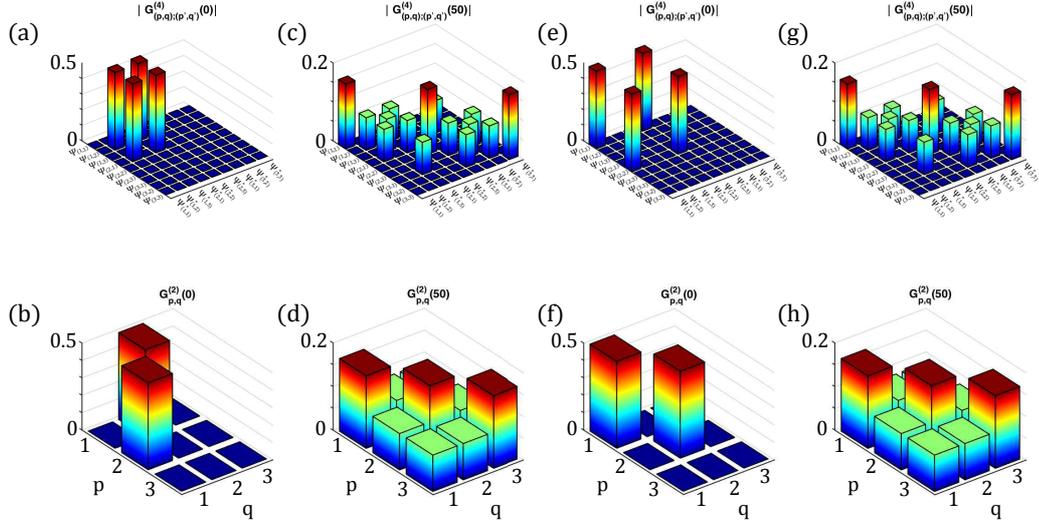}
\caption{(Top row) Initial $\left(\tilde{G}^{(4)}_{(p,q);(p',q')}(0)\right)$ and steady states $\left(\tilde{G}^{(4)}_{(p,q);(p',q')}(50)\right)$ for two-particle separable (a) and entangled (e) states. From (c) and (g) it is clearly seen that both steady states are identical and they retain the indistinguishable off-diagonal terms $\tilde{G}^{(4)}_{(p,q);(q,p)}(50)$. (Bottom row) Two-point correlation matrices obtained from the diagonal elements of (a), (c), (e), and (g), respectively. In the diagonals of (d) and (h) one can see that the bunching terms, $\tilde{G}^{(2)}_{(1,1)}=\tilde{G}^{(2)}_{(2,2)}=\tilde{G}^{(2)}_{(3,3)}$, show the highest probability, while the antibunching terms, $\tilde{G}^{(2)}_{(1,2)}=\tilde{G}^{(2)}_{(1,3)}=\tilde{G}^{(2)}_{(2,3)}$, have the second highest ones.}
\label{Fig3}
\end{figure}
To exemplify these effects, we explore the dynamics of separable i), entangled ii), classically correlated iii), and two distinguishable particles iv) as described in the previous section. Under such excitations, numerical integration of Eq.~\eqref{eq:average} renders the mean four-point correlation matrices displayed in Figs. (\ref{Fig3}) and (\ref{Fig4}).
For all simulations we assume that the site energies randomly change in the interval $\delta t=1$ (correlation time) and they obey a Gaussian distribution with variance $\sigma=2$ and mean values $\epsilon_{1}=\epsilon_{2}=\epsilon_{3}=1$. The dephasing rates are estimated using the relation $\g=\frac{\sigma^{2}\delta t}{2} =2$ \cite{Laing2008}. 
These numerical results clearly demonstrate that separable Fig.~(\ref{Fig3}.a), entangled Fig.~(\ref{Fig3}.e), and classically correlated Fig.~(\ref{Fig4}.a) bosons evolve into identical steady states, as shown in Figs.~(\ref{Fig3}.c), (\ref{Fig3}.g),  and (\ref{Fig4}.c), respectively. The main difference to note in comparison with the noiseless system from our previous section, is that in the present three cases, i), ii), and iii), the bunching correlation terms become equal, including those involving the third (lowest-coupled) site. Consequently, we can certainly state that dynamic disorder strengthens interactions between sites that are otherwise uncoupled or weakly coupled. Moreover, as implied by the first term on the right-hand side of Eq. \eqref{eq:average}, we observe that indistinguishable off-diagonal terms $\tilde{G}^{(4)}_{(p,q);(q,p)}$ remain immune to the impact of disorder, see off-diagonal elements in Figs.~(\ref{Fig3}.c), (\ref{Fig3}.g), and (\ref{Fig4}.c).\\  
\begin{figure}[t!]
\includegraphics[width=\linewidth]{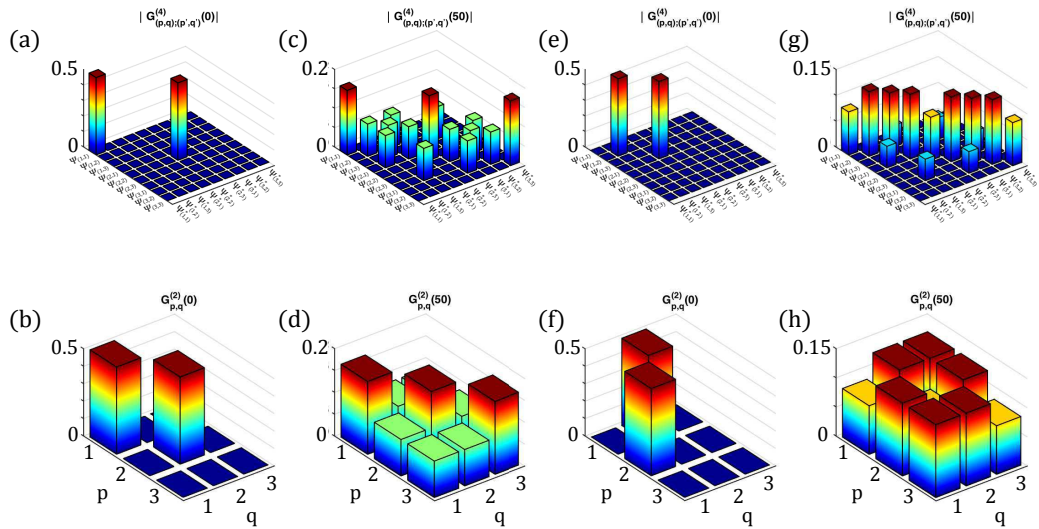}
\caption{(Top row) (a) and (e) depict the initial $\left(\tilde{G}^{(4)}_{(p,q);(p',q')}(0)\right)$ states for two-particle classically correlated and two distinguishable particles, respectively. (c) and (g) show the corresponding evolved states. Notice that the steady state for classically correlated particles  (c) is identical to the steady state obtained for separable and entangled particles. For distinguishable particles the state remains incoherent during evolution as shown in (g). From the $\tilde{G}^{(2)}_{(p,q)}$ shown in (d) and (h), we see for the classically correlated state the bunching terms exhibit the highest probabilities, while for the incoherent state the antibunching are the highest.}
\label{Fig4}
\end{figure}
At this point it is worth underlining that coherences of the type $\tilde{G}^{(4)}_{(p,q);(q,p)}$ arise by virtue of the wavefunction symmetrization which accounts for indistinguishability and exchange statistics of the particles \cite{Feynman1965}. However, we must point out that there is an ongoing debate regarding the observability or physical significance of correlations due to symmetrization \cite{Benatti2014, Reusch2015}. Indeed, the issue arises because correlations of the type $\left\langle\PP_{p,q}\PP^{*}_{q,p}\right\rangle$ represent superpositions of two-particle states where the only difference is the order of the particles. Formally, such coherences do not represent manipulable superpositions, but they are telltale of the fundamental particle indistinguishability. Consequently, the presence of such \textit{coherences} in the steady state imply that the particles retain their capability to interfere in experiments of the Hong-Ou-Mandel type \cite{Killoran2014}. Moreover, it has been shown that interferences stemming from the exchange symmetry can be made accessible  \cite{Cavalcanti2007}.
More precisely, if those states were to be used for exciting an external interferometer, to some extent they will show some interference as they have the necessary element to interfere, namely particle indistinguishablity \cite{Mandel1991, Zou1991, Saleh2000}. Nevertheless, in the present systems such states do not interfere since they form the parts of a stationary state (steady state). \\
To obtain a better understanding on the role of particle indistinguishability, we compare the correlation patterns generated by a mixed state that exhibits the strongest classical correlations where the particles are still indistinguishable, namely state iii), and an incoherent state where the particles are distinguishable, i.e., state iv). As stated above, initial state iii) gives rise to a steady state identical to the ones obtained for the coherent cases i) and ii), see Fig.~(\ref{Fig4}.c). Conversely, for distinguishable particles the state remains incoherent as elucidated by the diagonal elements in Fig.~(\ref{Fig4}.g). Therefore, if dynamic disorder destroys particle indistinguishability, in all cases one would expect to observe incoherent correlation patters similar to the one obtained for  distinguishable particles, see Fig.~(\ref{Fig4}.g). However, this is not the case as indistinguishable particles exhibit very different correlation patterns where the bunching terms $G^{(2)}_{(1,1)}=G^{(2)}_{(2,2)}=G^{(2)}_{(3,3)}$ are the highest ones, while the anti-bunching terms $G^{(2)}_{(1,2)}=G^{(2)}_{(1,3)}=G^{(2)}_{(2,3)}$ appear to be the second highest probabilities, Fig.~(\ref{Fig4}.d). On the contrary, for distinguishable particles the anti-bunching is the most probable event to occur and bunching is the second highest contribution, Fig.~(\ref{Fig4}.h). 
\section{IV. Summary and conclusions}
In this work, we have provided a model for describing two-particle quantum correlations in dynamically disordered tight-binding networks. Using this model, we have shown that dynamic disorder (or noise) creates new pathways through which multiple excitations can propagate, that is, it strengthens interactions between sites that are otherwise uncoupled or weakly coupled. Moreover, we have demonstrated that correlation elements accounting for particle indistinguishability are immune to the impact of noise. Our results may help elucidating the role of particle distinguishability to preserve quantum coherence and entanglement propagating through complex dynamically-disordered systems.

\appendix*
\section{APPENDIX:}
\setcounter{equation}{0}
The procedure to derive the equation of motion for the two-particle four-point correlation function Eq.~\eqref{eq:average} is given in this Appendix. 
We start by writing the expression for the time-evolution of the evolution operator's elements as
\begin{eqnarray}
\d U_{q,n} & = & \left(\im\epsilon_{q}U_{q,n}+\im\sum_{r=1}\kappa_{r,q}U_{r,n}+\im\phi_{q}(t)U_{q,n}\right)\d t, \\
\d U_{p,m} & = & \left(\im\epsilon_{p}U_{p,m}+\im\sum_{r=1}\kappa_{r,p}U_{r,m}+\im\phi_{p}(t)U_{p,m}\right)\d t.
\end{eqnarray}
We introduce the Wiener increments
\begin{eqnarray}
\d W_{p} & = & \frac{\phi_{p}(t)}{\sqrt{\gamma_{p}}}\d t, \\
\d W_{q} & = & \frac{\phi_{q}(t)}{\sqrt{\gamma_{q}}}\d t,
\end{eqnarray}
which upon substitution into Eqs. (A.1) and (A.2), respectively, yield
\begin{eqnarray}
\d U_{q,n} & = & \left(\im\epsilon_{q}U_{q,n}+\im\sum_{r=1}\kappa_{r,q}U_{r,n}\right)\d t+\im\sqrt{\gamma_{q}}U_{q,n}dW_{q}, \\
\d U_{p,m} & = & \left(\im\epsilon_{p}U_{p,m}+\im\sum_{r=1}\kappa_{r,p}U_{r,m}\right)\d t+\im\sqrt{\gamma_{p}}U_{p,m}dW_{p}.
\end{eqnarray}
Since we work in the It\^{o}'s calculus framework, we write these equations in their It\^{o} form \cite{Eisfeld2012}
\begin{eqnarray}
\d U_{q,n} & = & \left(\im\epsilon_{q}U_{q,n}+\im\sum_{r=1}\kappa_{r,q}U_{r,n}-\frac{1}{2}\gamma_{q}U_{q,n}\right)\d t+\im\sqrt{\gamma_{q}}U_{q,n}dW_{q},\\
\d U_{p,m} & = & \left(\im\epsilon_{p}U_{p,m}+\im\sum_{r=1}\kappa_{r,p}U_{r,m}-\frac{1}{2}\gamma_{p}U_{p,m}\right)\d t+\im\sqrt{\gamma_{p}}U_{p,m}dW_{p}.
\end{eqnarray}
If we take the product of $\d\left(U_{q,n}U_{p,m}\right)$ up to first order in $\d t$, we obtain
\begin{equation}
\begin{split}
\d \left(U_{q,n}U_{p,m}\right) & = \left[\im\left(\epsilon_{p}+\epsilon_{q}\right)U_{q,n}U_{p,m}+\im\sum_{r=1}\kappa_{r,p}U_{q,n}U_{r,m}\right.\\
&  \left. + \im\sum_{r=1}\kappa_{r,q}U_{r,n}U_{p,m}-\frac{1}{2}\left(\gamma_{p}+\gamma_{q}\right)U_{q,n}U_{p,m}\right]\d t\\
 &   +\im\left(\sqrt{\gamma_{p}}\d W_{p}+\sqrt{\gamma_{q}}\d W_{q}\right)U_{q,n} U_{p,m} -\sqrt{\gamma_{p}\gamma_{q}}U_{q,n}U_{p,m}\d W_{p}\d W_{q}.
 \end{split}
\end{equation}
Similarly, we have
\begin{equation}
\begin{split}
\d\left(U_{q,m}U_{p,n}\right) & =  \left[\im\left(\epsilon_{p}+\epsilon_{q}\right)U_{q,m}U_{p,n}+i\sum_{r=1}\kappa_{r,p}U_{q,m}U_{r,n}\right.\\
&  \left. +\im\sum_{r=1}\kappa_{r,q}U_{r,m}U_{p,n}-\frac{1}{2}\left(\gamma_{p}+\gamma_{q}\right)U_{q,m}U_{p,n}\right]\d t\\
 &   +\im\left(\sqrt{\gamma_{p}}dW_{p}+\sqrt{\gamma_{q}}\d W_{q}\right)U_{q,m}U_{p,n}-\sqrt{\gamma_{p}\gamma_{q}}U_{q,m}U_{p,n}\d W_{p}\d W_{q}
\end{split}
\end{equation}
Thus, adding these two contributions and following the It\^{o}'s product rule, $\d\left(U_{p,n}U_{q,m}\right)=\d\left(U_{p,n}\right)U_{q,m}+U_{p,n}\d\left(U_{q,m}\right)+\d\left(U_{p,n}\right)\d\left(U_{q,m}\right)$,  we obtain the expression
\begin{equation}
\begin{split}
\d\left(U_{p,n}U_{q,m}+U_{q,n}U_{p,m}\right) & =  \im\left(\epsilon_{p}+\epsilon_{q}\right)\left(U_{p,n}U_{q,m}+U_{q,n}U_{p,m}\right)\d t\\
&+ \im\sum_{r=1}\kappa_{r,q}\left(U_{p,n}U_{r,m}+U_{p,m}U_{r,n}\right)\d t\\
 & +  \im\sum_{r=1}\kappa_{r,p}\left(U_{r,n}U_{q,m}+U_{r,m}U_{q,n}\right)\d t\\
& -\frac{1}{2}\left(\gamma_{p}+\gamma_{q}\right)\left(U_{p,n}U_{q,m}+U_{q,n}U_{p,m}\right)\d t\\
 & +  \im\left(\sqrt{\gamma_{p}}\d W_{p}+\sqrt{\gamma_{q}}\d W_{q}\right)\left(U_{p,n}U_{q,m}+U_{q,n}U_{p,m}\right)\\
 & -  \sqrt{\gamma_{p}\gamma_{q}}\left(U_{p,n}U_{q,m}+U_{q,n}U_{p,m}\right)\d W_{p}\d W_{q}.
 \end{split}
\end{equation}
Hence, by using the definition of $\Psi_{p,q}$ given in Eq.~\eqref{eq:2}, we obtain
\begin{equation}
\begin{split}\label{eq:dPsi}
\d\Psi_{p,q}=&\left[\im\left(\epsilon_{p}+\epsilon_{q}\right)\Psi_{p,q}+\im\sum_{r=1}\kappa_{r,q}\Psi_{p,r}+\im\sum_{r=1}\kappa_{r,p}\Psi_{r,q}-\frac{1}{2}\left(\gamma_{p}+\gamma_{q}\right)\Psi_{p,q}\right]\d t\\
& +\im\sqrt{\gamma_{p}}\Psi_{p,q}\d W_{p} + \im\sqrt{\gamma_{q}}\Psi_{p,q}\d W_{q}-\sqrt{\gamma_{p}\gamma_{q}}\Psi_{p,q}\d W_{p}\d W_{q}.
\end{split}
\end{equation}
Now, by using the It\^{o}'s product rule $\d\left(\Psi_{p,q}\Psi_{p',q'}^{*}\right)=\Psi_{p,q}\d\left(\Psi_{p',q'}^{*}\right)+\Psi_{p',q'}^{*}\d\left(\Psi_{p,q}\right)+\d\left(\Psi_{p,q}\right)\d\left(\Psi_{p',q'}^{*}\right)$, along with Eq.~\eqref{eq:dPsi} and its complex conjugate, we obtain
\begin{eqnarray*}
\d\left(\Psi_{p,q}\Psi_{p',q'}^{*}\right)  & = & \left[-\im(\epsilon_{p'}+\epsilon_{q'})\Psi_{p,q}\Psi_{p',q'}^{*}-\im\sum_{r=1}\kappa_{r,q'}\Psi_{p,q}\Psi_{p',r}^{*}\right.\\
 &  &\;\;\;\;\;\;\;\;  \left. -\im\sum_{r=1}\kappa_{r,p'}\Psi_{p,q}\Psi_{r,q'}^{*}-\frac{1}{2}(\gamma_{p'}+\gamma_{q'})\Psi_{p,q}\Psi_{p',q'}^{*}\right]\d t \\
 &  & +\left[\im(\epsilon_{p}+\epsilon_{q})\Psi_{p,q}\Psi_{p',q'}^{*}+\im\sum_{r=1}\kappa_{r,q}\Psi_{p,r}\Psi_{p',q'}^{*}\right.\\
 &  &\;\;\;\;\;\;\;\;\;  +\left. \im\sum_{r=1}\kappa_{r,p}\Psi_{r,q}\Psi_{p',q'}^{*}-\frac{1}{2}(\gamma_{p}+\gamma_{q})\Psi_{p,q}\Psi_{p',q'}^{*}\right]\d t\\
&  & -\sqrt{\gamma_{p'}\gamma_{q'}}\Psi_{p,q}\Psi_{p',q'}^{*}\d W_{p'}\d W_{q'} -\sqrt{\gamma_{p}\gamma_{q}}\Psi_{p,q}\Psi_{p',q'}^{*}\d W_{p}\d W_{q}\\
 &  & +\sqrt{\gamma_{p}\gamma_{p'}}\Psi_{p,q}\Psi_{p',q'}^{*}\d W_{p}\d W_{p'}+\sqrt{\gamma_{q}\gamma_{q'}}\Psi_{p,q}\Psi_{p',q'}^{*}\d W_{q}\d W_{q'}\\
 &  & +\sqrt{\gamma_{p}\gamma_{q'}}\Psi_{p,q}\Psi_{p',q'}^{*}\d W_{p}\d W_{q'} + \sqrt{\gamma_{q}\gamma_{p'}}\Psi_{p,q}\Psi_{p',q'}^{*}\d W_{q}\d W_{p'}.
\end{eqnarray*}
Taking the average, we finally obtain
\begin{eqnarray}
\frac{\d\left\langle \Psi_{p,q}\Psi_{p',q'}^{*}\right\rangle }{\d t} & = & \Bigg[\im\left(\epsilon_{p}+\epsilon_{q}-\epsilon_{p'}-\epsilon_{q'}\right)-\frac{1}{2}\left(\gamma_{p}+\gamma_{q}+\gamma_{p'}+\gamma_{q'}\right) \nonumber \\
&  &\;\;\;\;\; \left. +\sqrt{\gamma_{p}\gamma_{p'}}\delta_{pp'}+\sqrt{\gamma_{q}\gamma_{q'}}\delta_{qq'}+\sqrt{\gamma_{p}\gamma_{q'}}\delta_{pq'}+\sqrt{\gamma_{q}\gamma_{p'}}\delta_{q,p'}\right. \nonumber \\
&  &\;\;\;\;\;\;\;\;\;\;\;\;\;\;\; -\sqrt{\gamma_{p}\gamma_{q}}\delta_{p,q}-\sqrt{\gamma_{p'}\gamma_{q'}}\delta_{p',q'}\Bigg]\left\langle \Psi_{p,q}\Psi_{p',q'}^{*}\right\rangle \nonumber \\
 &  & +\im\Bigg[\sum_{r=1}\kappa_{r,q}\left\langle \Psi_{p,r}\Psi_{p',q'}^{*}\right\rangle +\sum_{r=1}\kappa_{r,p}\left\langle \Psi_{r,q}\Psi_{p',q'}^{*}\right\rangle \nonumber \\
 &  & \;\;\;\;\;\;\;\;\;\;\;\;\;\;\;  -\sum_{r=1}\kappa_{r,q'}\left\langle \Psi_{p,q}\Psi_{p',r}^{*}\right\rangle -\sum_{r=1}\kappa_{r,p'}\left\langle \Psi_{p,q}\Psi_{r,q'}^{*}\right\rangle \Bigg],
\end{eqnarray}
where $\delta_{p,q}$ is the Kronecker delta. After the substitution $\left\langle \Psi_{p,q}(t)\Psi_{p',q'}^{*}(t)\right\rangle=\tilde{G}^{(4)}_{(p,q);(p',q')}(t)$ Eq.~\eqref{eq:average} follows.
\section*{ACKNOWLEDGMENT}
A. P.-L and K. B. acknowledge financial support by the Deutsche Forschungsgemeinschaft (DFG) within project BU 1107/10-1 of the Priority Program SPP 1839 {\it Tailored Disorder}. H. M.-C. thanks the Alexander von Humboldt Foundation for support.

\end{document}